\documentclass[prb, twocolumn,amsmath,amssymb]{revtex4}
\usepackage{graphicx, color}
\usepackage{subfigure}
\usepackage{dcolumn}
\usepackage{bm}
\usepackage{mathrsfs}
\usepackage{color}
\usepackage{epstopdf}
\begin{document}
\title{Equilibrium angular momentum and edge current in Bose-condensed cold atom systems with $k$-space Berry curvature}
\author{Xiao-Hui Li$^{1}$}
\author{Ting-Pong Choy$^{1}$}
\author{Tai-Kai Ng$^{1,2}$}
\affiliation{$^{1}$ Department of Physics, Hong Kong University of Science and Technology, Clear Water Bay, Hong Kong, China}
\affiliation{$^{2}$ Hong Kong Academy for Gifted Education, Shatin, Hong Kong, China}
\date{\today}
\begin{abstract}
 In this paper we study the properties of cold bosons in a two-dimensional optical lattice system where Bose-condensation occurs at a momentum point $\mathbf{k}$ with non-zero $k$-space Berry curvature. By combining results from both analytic and numerical approaches, we show that the boson system carries non-universal, temperature dependent equilibrium angular momentum and edge current at low temperatures.
\end{abstract}
\maketitle
It is well known that Berry phase can induce profound effect on electronic system\cite{Chang1996Berry,xiao2010berry} and is closely related to non-trivial topological phenomena like the quantum Hall effect\cite{klitzing1980new,tsui1982two}, quantum spin Hall effect\cite{bernevig2006quantum} and quantum anomalous Hall effect\cite{laughlin1983anomalous,chang2013experimental}.  The topological invariant (Chern number) characterizing these systems is the integral of Berry curvature $\Omega(\mathbf{k})$ over the entire Brillioun zone\cite{Simon1983Holonomy}. The $k$-space Berry curvature is also predicted to have direct consequences on electronic transports\cite{Chang1995Hofstadter,xiao2010berry}. The Berry curvature can be induced by spin-orbit coupling (SOC) and several theoretical schemes to realize SOC in cold atom systems have been proposed in the last few years\cite{dudarev2004spin,osterloh2005cold,ho2011bose,campbell2011realistic,Zhai2012Super,liu2014realization}, with some of them being implemented in optical lattice experimentally\cite{lin2009synthetic,lin2011spin,galitski2013spin}. 
These efforts open a new route for scientists to explore plausible Berry curvature-induced exotic phases in cold atom systems. For instance, realization of the Haldane model\cite{jotzu2014experimental}, Majorana fermions in ultracold gas\cite{sato2009non,jiang2011majorana}, the chiral interacting bosonic superfluid\cite{li2014chiral} and the chiral Meissner currents in bosonic ladders\cite{atala2014observation}.

 The combined effect of Bose-condensation and $k$-space Berry curvature is a natural direction of research in this new area. Recently, several related works have been done within the semiclassical approximation\cite{Price2013Collective,yao2008berry}. In this paper, starting from the Gross-Pitaevskii (GP) equation, we study a simple model of weakly interacting bosons moving in a lattice with nonzero $k$-space Berry curvature at the momentum point where Bose-condensation occurs. By combining results from both analytic and numerical approaches, we show that non-universal angular momentum and edge current exist in the system at low temperatures. The angular momentum has both bulk and edge contributions. Furthermore, gapless bulk and chiral edge excitations coexist which lead to the modification of these entities at non-zero temperatures.

We start with a continuum model of spinless bosons in a two-dimensional optical lattice trapped in an external potential $V(\mathbf{R})$. The system is described by the GP equation\cite{pethick2002bose}  $i\hbar\frac{\partial \Psi}{\partial t}= \hat{H} \Psi$  where
\begin{align}
\label{h}
\hat{H} & = -{\hbar^2\over2m}\nabla^2_{\mathbf{r}}+V_{\rm eff}(\mathbf{r}+\mathbf{A}_p(\mathbf{p})),\\
V_{\rm eff}(\mathbf{R})&=  V(\mathbf{R})+U\rho(\mathbf{R})-\mu. \nonumber
\end{align}
$V_{\rm eff}(\bm R)$ is the effective Hamiltonian acting on the boson condensate wavefunction $\Psi$, $\rho(\mathbf{R})=\Psi^+(\mathbf{R})\Psi(\mathbf{R})$ is the boson density and $U$ is the interaction between bosons. The optical lattice leads to non-vanishing $k$-space Berry curvature at the momentum point $\mathbf{p}_0$ (which we shall set to be zero in the following) where boson condenses. As a result the physical position and momentum $(\mathbf{R},\mathbf{p})$ are not equivalent to the conjugate pair $(\mathbf{r},\mathbf{p}=-i\hbar\nabla_{\mathbf{r}})$ anymore. Instead the physical position is related to $\mathbf{r}$ by $\mathbf{R} = \mathbf{r}+\mathbf{A}_p(\mathbf{p})$\cite{yao2008berry} where $\mathbf{A}_p(\mathbf{p})$ is the $k$-space Berry connection and $\Omega(\mathbf{p}) = \nabla_{\mathbf{p}}\times \mathbf{A}_p(\mathbf{p})$ is the Berry curvature at the momentum $\mathbf{p}$-point. We note that the existence of $k$-space Berry curvature at the Bose-condensation point implies that Time-Reversal Symmetry is broken in the system.

The kinetic energy term in\ (\ref{h}) should be understood as the expansion of the band energy of the bosons in optical lattice around band minima $\mathbf{p}_0$ to second order in $\mathbf{p}=-i\hbar\nabla_{\mathbf{r}}$. To be consistent with this expansion we also expand the potential energy term to second order in the Berry curvature term $\mathbf{A}_p$, obtaining
 \[
  V_{\rm eff}(\mathbf{R})\sim \left(I+\mathbf{A}_p(\mathbf{p})\cdot\nabla_{\mathbf{r}} +{1\over2}(\mathbf{A}_p(\mathbf{p})\cdot\nabla_{\mathbf{r}})^2\right) V_{\rm eff}(\mathbf{r}). \]

 The equation can be further simplified if we approximate $\Omega(\mathbf{p})\sim\Omega(\mathbf{p}=0)$, consistent with keeping terms up to second order in $\mathbf{p}$ in the GP equation. In this case we may write $\mathbf{A}_p(\mathbf{p})=-{1\over2}\mathbf{p}\times\mathbf{\Omega}$ (radial gauge), where $\mathbf{\Omega}=\Omega(\mathbf{p}=0)\hat{z}$. We see that $\mathbf{A}_p(\mathbf{p})$ is of first order in $\mathbf{p}$ itself thus to second order in $\nabla_{\mathbf{r}}$, we only need to keep the first order term $\mathbf{A}_p(\mathbf{p})\cdot\nabla_{\mathbf{r}}$ in the expansion of $V_{\rm eff}(\mathbf{R})$, obtaining
 \begin{subequations}
 \label{h2}
 \begin{equation}
\hat{H}\rightarrow 
{\mathbf{p}^2\over2m}+{1\over2}(\mathbf{A}(\mathbf{r})\cdot\mathbf{p}+\mathbf{p}\cdot\mathbf{A}(\mathbf{r}))+V_{\rm eff}(\mathbf{r})
-\mu
\end{equation}
where
\begin{equation}
\mathbf{A}(\mathbf{r})=-{1\over2}(\mathbf{\Omega}\times\nabla V_{\rm eff}(\mathbf{r})),
\end{equation}
\end{subequations}
  i.e., the $k$-space Berry curvature introduces an effective {\em real space} vector field $\mathbf{A}(\mathbf{r})$ coupled to bosons in the presence of $\nabla V_{\rm eff}(\mathbf{r})\neq0$. 
In the following we study the predictions of this effective GP equation.

We first distinguish the canonical velocity operators $\mathbf{v}_c$ and the physical velocity operator $\mathbf{v}$ given by
\begin{subequations}
\begin{equation}
\label{current}
\mathbf{v}={d\mathbf{R}\over dt}={d\mathbf{r}\over dt}+ {d\mathbf{A}_p(\mathbf{p})\over dt}=\mathbf{v}_c+ {d\mathbf{A}_p(\mathbf{p})\over dt}.
\end{equation}
It is straightforward to show that
\begin{equation}
\label{vc}
\mathbf{v}_c={1\over i\hbar}[\mathbf{r},\hat{H}]={\mathbf{p}\over m}+\mathbf{A}(\mathbf{r})
\end{equation}
and
   \begin{equation}
  \label{v}
   \mathbf{v}={\mathbf{v}}_c-{1\over2i\hbar}[\mathbf{p},\hat{H}]\times \mathbf{\Omega}=\mathbf{v}_c+\mathbf{A}(\mathbf{r})
  \end{equation}
   \end{subequations}
    to second order in $\nabla_{\mathbf{r}}$. 

Correspondingly the real space angular momentum operator is given by
\begin{eqnarray}
\label{amomentum}
\mathbf{L} & = & \mathbf{R}\times(m{\mathbf{v}})=\left(\mathbf{r}-{1\over2}(\mathbf{p}\times\mathbf{\Omega})
\right)\times (m\mathbf{v}) \\ \nonumber
 & = & \mathbf{L}_c-{\mathbf{\Omega}\over2}(\mathbf{p}\cdot m\mathbf{v})
\end{eqnarray}
 where $\mathbf{L}_c=m\mathbf{r}\times{\mathbf{v}}$ is the orbital angular momentum. We find that in additional to $\mathbf{L}_c$, the presence of non-zero Berry curvature leads to an intrinsic angular momentum $\sim\mathbf{\Omega}(\mathbf{p}\cdot m\mathbf{v})$, independent of microscopic details. This is consistent with the finding by Chang and Niu\cite{Chang1996Berry}, where they show that besides center of mass motion, a wave packet possesses in general also a self-rotation around its center of mass in the presence of $k$-space Berry curvature. The first and second terms in Eq.\ (\ref{amomentum}) describes the orbital and self-rotation motion of the wave packet, respectively.

We now apply these results to the boson system. We start with the local equilibrium (Thomas-Fermi) approximation (LEA) for the ground state where the solution of the GP equation\ (\ref{h2}) is given approximately by
 \begin{equation}
  \label{tf}
\Psi_0(\mathbf{r})=\sqrt{\rho_0(\mathbf{r})}=\left({\mu-V(\mathbf{r})\over U}\right)^{1\over2},
\end{equation}
for $\mu>V(\mathbf{r})$ and $\Psi_0(\mathbf{r})=0$ for $\mu<V(\mathbf{r})$. Notice that $V_{\rm eff}(\mathbf{r})=V(\mathbf{r})+U\rho(\mathbf{r})-\mu=0$ at places where $\Psi_0(\mathbf{r})\neq0$ and $\mathbf{A}(\mathbf{r})=0$ has no effect on $\Psi_0(\mathbf{r})$ in this approximation.

To see the effects of $k$-space Berry curvature we consider systems with rotational symmetry (i.e. $V(\mathbf{r})=V(r)$ where $r=|\mathbf{r}|$) and examining the angular momentum carried by the ground state wave-function. We obtain in LEA
\begin{eqnarray}
\label{amomentum}
\mathbf{L} & = & 2\pi\int rdr\Psi^+_0(r)[\mathbf{L}_c-{\mathbf{\Omega}\over2}(\mathbf{p}\cdot m\mathbf{v})]\Psi_0(r)
\\ \nonumber
& = & -\pi\mathbf{\Omega}\int^{r_c}_0 rdr\hbar^2({\partial\Psi_0(r)\over\partial r})^2
\end{eqnarray}
 where $\mu=V(r_c)$. Notice that $\mathbf{A}(\mathbf{r})$ has no effect on $\Psi_0(r)$ in LEA and $\mathbf{L}_c=0$. As a result there is only {\em self-rotation contribution} to angular momentum in LEA.


This expression for angular momentum has an interesting consequence for potential energy of form $V(r)=V(r/r_c)$ such that $r_c$ measures the size of the system. (Notice that the Berry curvature has no effect on $\Psi_0$ in LEA. As a result, $r=R$ measures the physical positions in the system). In this case the condensate wavefunction\ (\ref{tf}) satisfies
\begin{align}
N=2\pi\int^{r_c}_0 r dr|\psi_0(r)|^2=(\pi r_c^2)n,  \nonumber
\end{align}
where $N$ is the total boson number in the system and $n=2\int_0^1xdx|\psi_0(x)|^2$ is the boson density, $x=r/r_c$. However the total angular momentum is given by
\begin{align}
\mathbf{L}=-\hbar^2\pi\mathbf{\Omega}\int^{1}_0 xdx({\partial\Psi_0(x)\over\partial x})^2  \nonumber
\end{align}
is independent of the size of the system although $N$ scales as the area $\pi r_c^2$ at fixed density.

 It should be cautioned that the absence of orbital contribution to angular momentum is a special feature of LEA which breaks down at the edge of the system. We shall see later that in general an orbital edge motion also contributes to the angular momentum of the system. The edge angular momentum $L_c$ is of order $(2\pi r_c)\times r_c\times j_d$, where $j_d$ is the edge current ($j_d=0$ in LEA). The total angular momentum $L$ scales as $L\sim a+br_c^2$ in the presence of both self-rotation and edge (orbital) contributions.

%
%

\begin{figure}[htbp]
\centering
\includegraphics[width=2.6in]{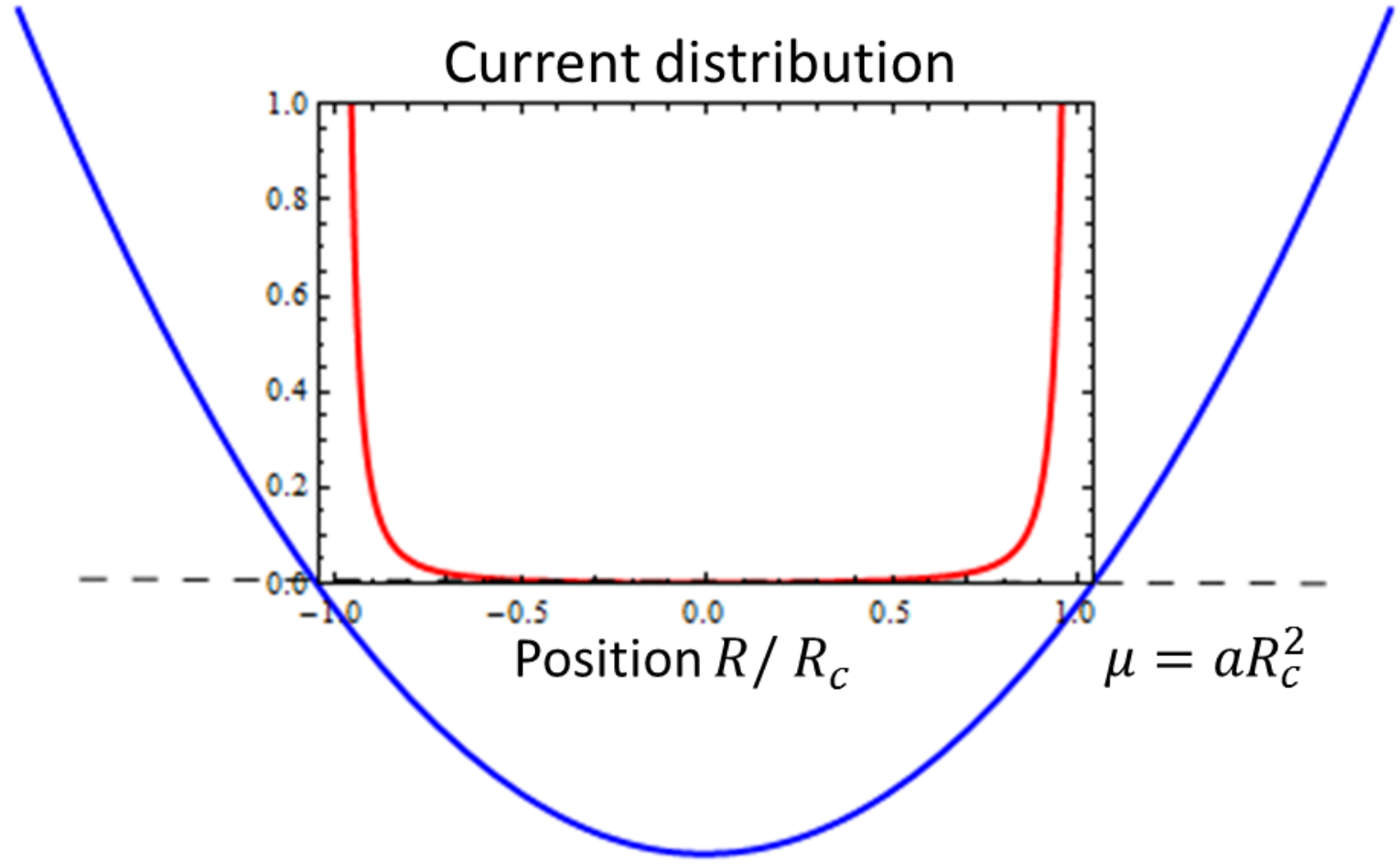}
\caption{The distribution of current from the analytic result in a harmonic trap. $\mu$ is the chemical potential and $r_c$ is the physical boundary of the system. One can see LEA fails at the physical boundary. }\label{Fig:Current_analytic}
\end{figure}

The total angular momentum Eq.\eqref{amomentum} suggests that the Bose-condensate carries a bulk angular momentum density
\begin{equation} \label{adensity}
\mathbf{m}(\mathbf{r})=-{\hbar^2\mathbf{\Omega}\over2}({\partial\Psi_0(r)\over\partial r})^2
\end{equation}
which can be interpreted as coming from a rotating mass current given by $\mathbf{j}(\mathbf{r})=\frac{1}{m}\nabla_{\mathbf{r}}\times\mathbf{m}(\mathbf{r})$. Using Eq.\ (\ref{tf}) for $\Psi_0$, we find that LEA predicts that this rotating current is given by $\mathbf{j}(\mathbf{r})=j(r)\hat{\phi}$, where $\hat{\phi}$ is the azimuth angle in the two-dimensional plane and
\begin{align}
j(r)=&{\hbar^2\Omega\over8mU}{\partial\over\partial r}\left({1\over(\mu-V(r))}({\partial V(r)\over\partial r})^2\right).\nonumber
\end{align}
Notice that $j(r)$ depends strongly on microscopic details of the system and has no simple, {\em universal} character. The rotating current $j(r)$ as a function of $R=r$ is shown in Fig.\ref{Fig:Current_analytic} with a harmonic trapping potential $V(x)=ax^2$. The current $j(r)$ diverges as $R\rightarrow R_c$, reflecting the break down of LEA at the edge of the boson liquid. Nevertheless, the LEA suggests that circulating current exists in the ground state of a Boson system which Bose-condensate at a momentum point with non-zero $k$-space Berry curvature. The circulating current concentrates mainly at the edge of the sample, and is non-universal with structure depending on the details of the microscopic Hamiltonian.

To go beyond LEA we perform a numerical simulation on a tight-binding Hamiltonian with $k$-space Berry curvature. We consider interacting spin-$1/2$ bosons moving on a square lattice,
\begin{align}\label{TightBindingH}
H=&\sum_{\textbf{R},\textbf{d},\sigma} (-tb_{\textbf{R},\sigma}^\dagger b_{\textbf{R}+\textbf{d},\sigma}+ i\alpha \bm{\hat{z}} \cdot(\bm\sigma \times \textbf{d})  b_{\textbf{R},\sigma}^\dagger b_{\textbf{R}+\textbf{d},\sigma'}+h.c.) \nonumber\\
&+B_z\sum_{\textbf{R},\sigma}b_{\textbf{R},\sigma}^\dagger b_{\textbf{R},\sigma} \sigma_z+\sum_{\textbf{R},\sigma}U_\sigma n_{\textbf{R},\sigma}^2
\end{align}
where $t,\alpha,B_z,U$ are the hopping energy, Rashba SOC, Zeeman energy, the on-site repulsive interaction respectively, and $n_{\textbf{R}}=b_{\textbf{R},\sigma}^\dagger b_{\textbf{R},\sigma}$ is the boson density. We apply a self-consistent approach by replacing $Un_{\textbf{R}}^2$ by $U\langle n_{\textbf{R}}\rangle n_{\textbf{R}}$ where $\langle..\rangle$ is the ground state expectation value.

\begin{figure}[htbp]
\centering
\subfigure{
\includegraphics[width=1.5in]{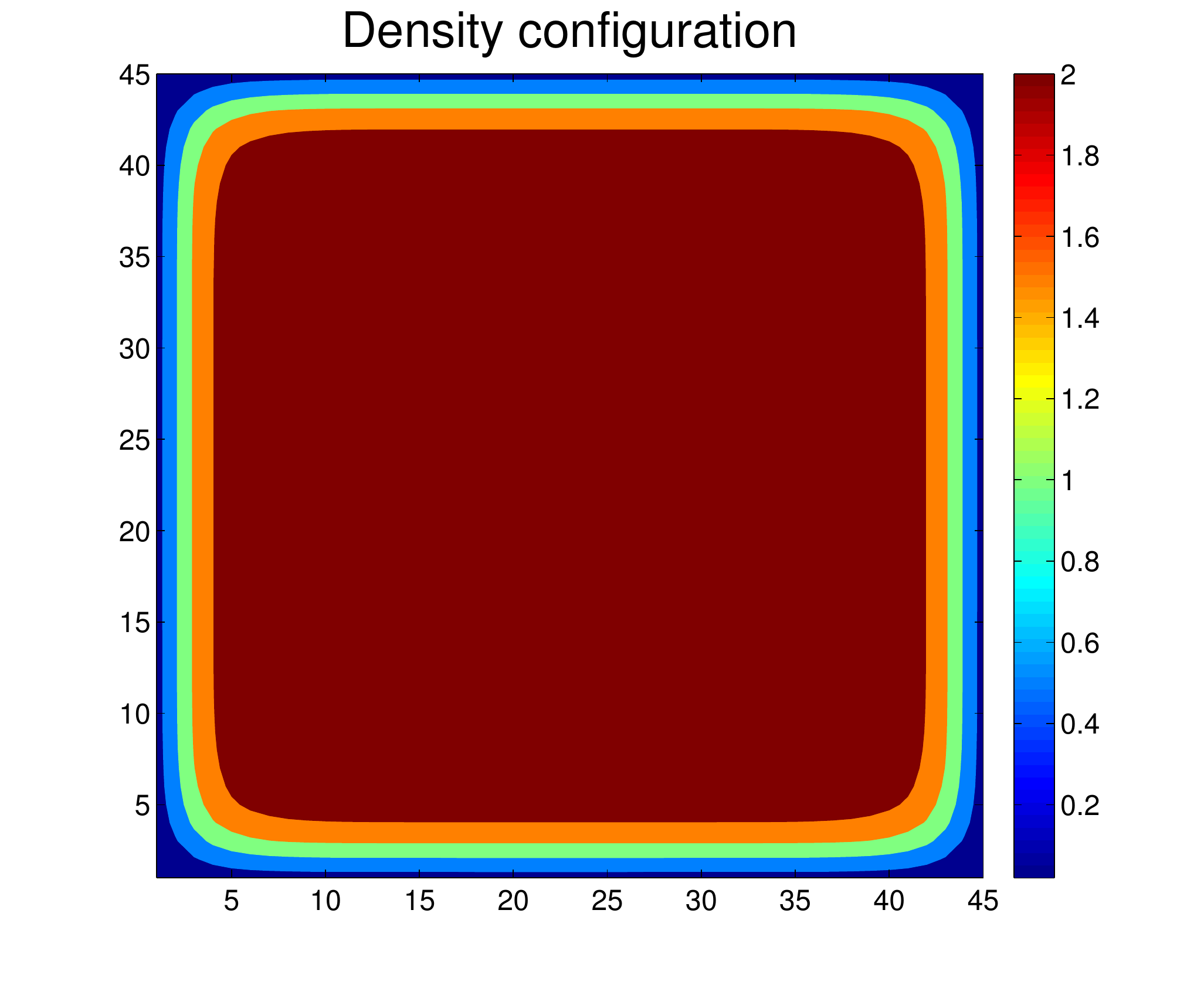}\label{Fig:Density}
}
\subfigure{
\includegraphics[width=1.4in]{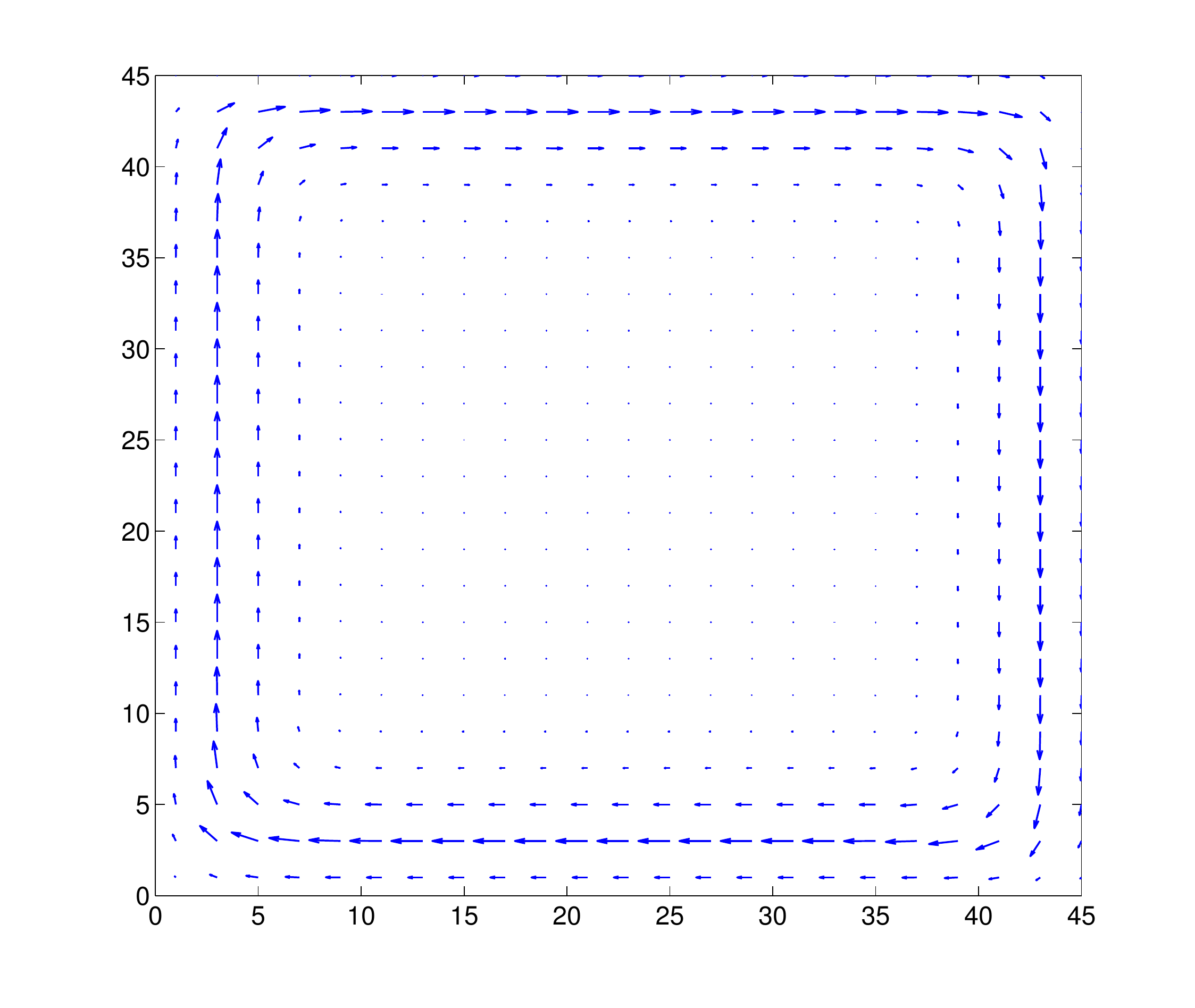}\label{Fig:currentflow}
}
\caption{The density configuration (Left) and the current flow (Right) in the self-consistent calculation. We take  $t=2$, $\alpha=0.3$, $B_z=1$, $U=0.2$ and $N=45$ in our calculation. The distribution of boson density is consistent with LEA except near the edge. The current is chiral and localized around the edge. } \label{Fig:density&current}
\end{figure}

The non-interacting single particle energy spectrum is given by $\mathscr{E}_{\pm}(\textbf{k})=-2t(\cos k_x+\cos k_y)\pm \sqrt{\alpha^2( \sin^2{k_x}+\sin^2 k_y)+B_z^2}$ . Expanding around $\textbf{k}=0$, we see that when $B_z>\frac{\alpha^2}{t}$, the spectrum has a single minimum $\mathscr{E}_{-}(\textbf{k})$ located at $\textbf{k}=0$. 
In our numerical simulation, we consider a weakly interacting boson gas in this regime where boson only condenses at the state $\mathscr{E}_{-}(\textbf{k}=0)$ and the ground state is affected solely by the Berry curvature at $\mathscr{E}_{-}(\textbf{k}\sim0)$. We consider a finite-size, $N\times N$ sites sample with open boundary and apply the optimal damping algorithm to determine the ground state self-consistently\cite{dion2007ground}.  The self-consistent ground state boson density distribution and current flow are shown in Fig.\ref{Fig:density&current}.
We can see that the boson density $\rho(x,y)$ is nearly uniform in the bulk of the sample and smoothly goes to zero at the edge. Indeed the chiral current appears which is not found in LEA. This current is mainly localized at the edge region where the density $\rho$ drops rapidly.

\begin{figure}[htbp]
 \centering
 \includegraphics[width=3.2in]{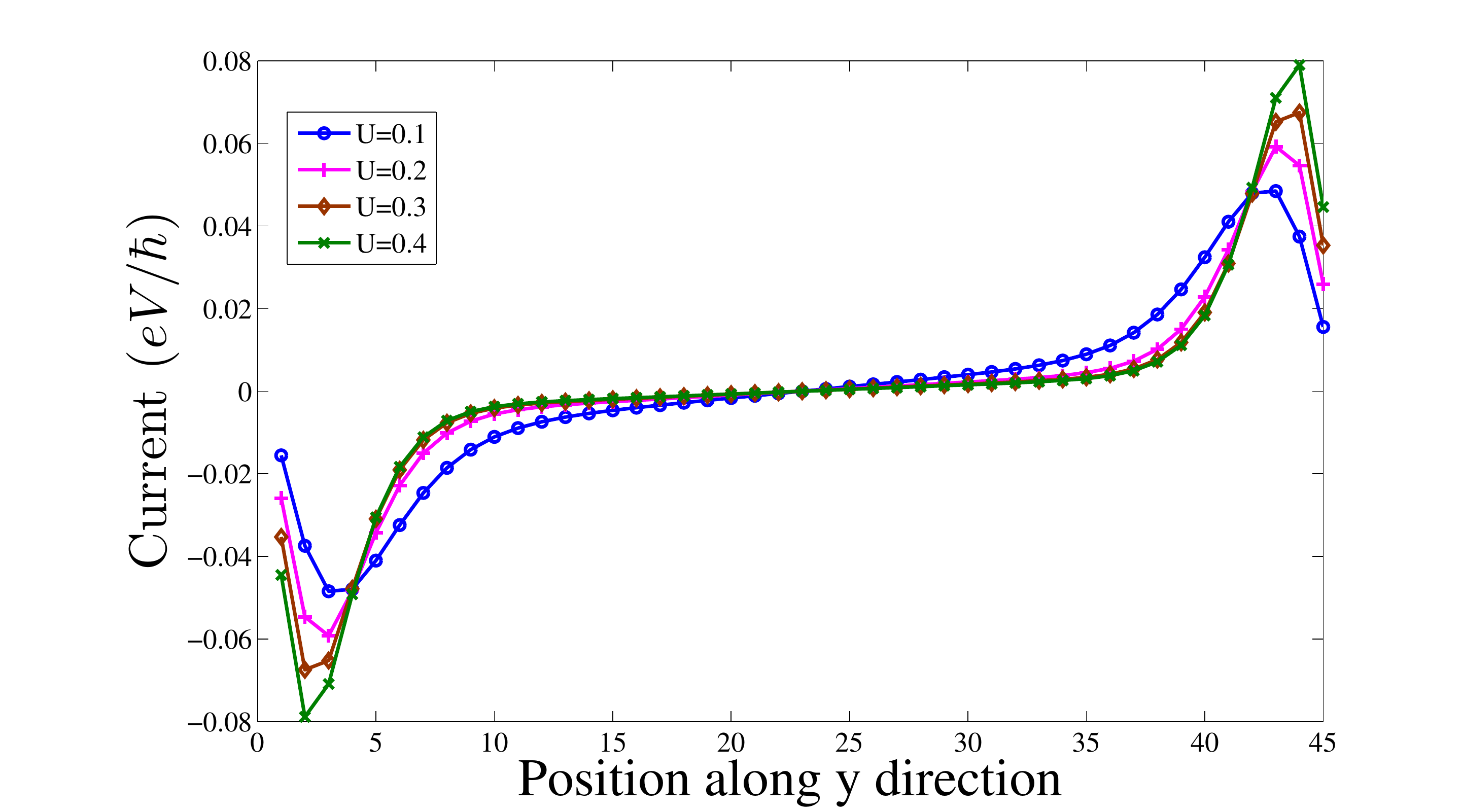}
 \caption{The charge current distribution in our model Eq.\eqref{TightBindingH} for several values of interaction strength $U$. The value of other parameters are the same as in Fig.\ref{Fig:density&current}. The current is mainly localized at the edge and the localization is strengthened by increasing $U$. }\label{Fig:CurrentDistribution}
 \end{figure}

 To look at the edge current more carefully we show the current $j_x(y)$ flowing in $x$ direction as a function of position $y$ at fixed $x=23$ for several different values of $U$ in Fig.\ref{Fig:CurrentDistribution}. We see that the edge region becomes smaller and sharper when the interacting parameter $U$ increases. We emphasize here that the chiral edge current we observed is qualitatively different from the ones in gapped topological systems :(1) the Chern number in our system is not required to be nonzero, since the edge current is related to Berry curvature at a local $k$-point only; (2) our system is a gappless system, owing to the spontaneous $U(1)$ symmetry breaking.

%

\begin{figure}[htbp]
\centering
\includegraphics[width=3.0in]{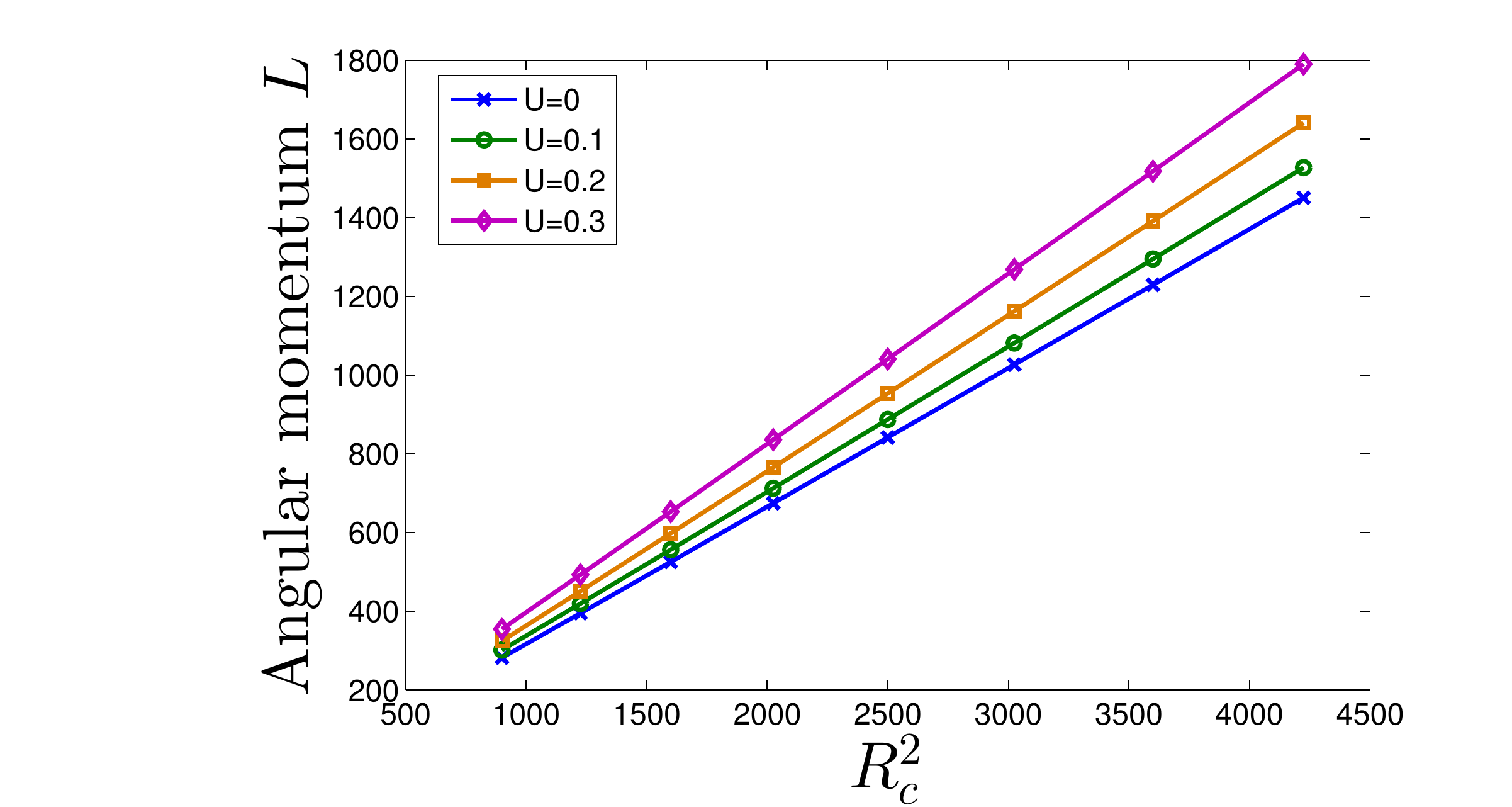}
\caption{The scaling relation between $L$ and size dimension $N\sim R_c$ for different interacting strength. We find that the total angular momentum scales as $L\sim a +bR_c^2$. Notice $b$ is a function of $U$.}\label{Fig:Lz-N}
\end{figure}

To examine the scaling relation between angular momentum and sample size $N\sim R_c$, we fix the average density $n=2$ per site in our simulation. The $L-N^2$ relation is shown in Fig.\ref{Fig:Lz-N} for several values of $U$. We see from Fig.\ref{Fig:Lz-N} that the total angular momentum scales $L\sim a+bR_c^2$, consistent with our analytical result. Notice $b$ depends on $U$, consistent with our expectation that $L$ is non-universal and depends on the microscopic Hamiltonian.

 Next we consider the effect of Berry curvature on the low energy excitations. For trapped bosons, there are bulk excitation modes and excitation modes localized near the edge of the boson cloud. The excitations can be obtained by writing $\Psi(\mathbf{r})=\sqrt{\rho_0(\mathbf{r})+\delta\rho(\mathbf{r},t)}e^{i\delta\theta(\mathbf{r}.t)}$ and solves the linearized GP equation\cite{pethick2002bose} for $\delta\rho$ and $\delta\theta$.

  Keeping terms to order $\nabla_{\mathbf{r}}^2$, we obtain after some straightforward algebra the well known hydrodynamic equations\cite{Price2013Collective}
  \begin{subequations}
  \label{hy}
\begin{equation}\label{Continuity_Equation}
{\partial{\delta\rho}\over\partial t}+\nabla \cdot(\rho_{0} \text{v}_c) =0
\end{equation}
 and
\begin{equation}\label{Acceleration}
 m{\partial \textbf{v}_c\over\partial t}=-\nabla (V(\textbf{r})+U\delta\rho)-(\frac{mU}{2}\nabla ({\partial\delta\rho\over\partial t})\times \mathbf{\Omega})
\end{equation}
\end{subequations}
 where $\mathbf{v}_c={\hbar\over m}\nabla\delta\theta+\mathbf{A}=d\mathbf{r}/dt$, in agreement with Eq.\ (\ref{v}).

At the bulk of the sample, $\nabla V=0$, $\rho_0$ is a constant and it is straightforward to obtain the usual Goldstone mode $\delta\rho(\mathbf{r},t)=\delta\rho e^{i(\omega t-\mathbf{q}.\mathbf{r})}$, with dispersion 
\begin{align}
\omega^2={\rho_0U\over m}q^2.\nonumber
\end{align}
Notice that the Berry curvature has no effect on the bulk dispersion.

The situation is rather different in the surface region. By approximating the potential in the surface region (in $x$-direction) as a linear function of the coordinate $V(\textbf{r})-\mu\sim Fx$ and assuming the LEA for $\rho_0$, we obtain for $q<<\delta^{-1}$, where $\delta\sim(\hbar^2/2mF)^{\frac{1}{3}}$,
\begin{equation}\label{Surface_density}
\delta\rho(x,y,t)=f(qx) e^{-q|x|\pm iqy}e^{-i\omega t}. (q>0)
\end{equation}
and $\pm q$ represents two different modes propagating along positive and negative $y$ direction, respectively.


The solution for Eq.\ ({\ref{hy}) is the Laguerre polynomials $L_n(2qx)$ with dispersion relation
\begin{equation}
\label{smode}
\omega^{\pm}_n(q)=\sqrt{(2n+1)\frac{Fq}{m}+\frac{\Omega^2F^2q^2}{16}}\pm\frac{\Omega Fq}{4}
\end{equation}
consistent with the result obtained by Price and Cooper\cite{Price2013Collective}. Note that for $\Omega\neq 0$, the excitation energies correspond to the left and right propagating modes are different.

The surface excitation results in a $T\neq 0$ contribution to the equilibrium surface current given by
\begin{equation}
j(T)=\sum_{n,q>0} \left( n_B(\omega^+_n(q))v_n^+(q)-n_B(\omega^-_n(q))v_n^-(q) \right)
\end{equation}
where $v_n^{\pm}(q)=\nabla_{q}\omega^{\pm}_n(q)$ is the group velocity and $n_B(\omega)$ is the Bose-Einstein distribution function.
Using the fact that $v^{+(-)}_n(q)=\frac{\partial \omega^{+(-)}_n(q)}{\partial q}$ in 1D, we obtain
\begin{align}\label{FiTcurrent}
j_S(T)&=\frac{2\pi r_c}{2\pi} \left( \int_{0}^{\omega_{\Lambda}^{+}}d\omega^+(q)  n_B(\omega^+_q)-\int_{0}^{\omega_{\Lambda}^{-}} d\omega^-(q)   n_B(\omega^-_q) \right),
\nonumber\\
&\sim r_c (k_BT) \left( e^{-{\omega^-_{\Lambda}\over k_BT}}-e^{-{\omega^+_{\Lambda}\over k_BT}} \right),
\end{align}
where $j_S(T)$ depends on the high momentum cutoff. Here we only consider $n=0$ contribution but similar conclusion can be drawn for $n\neq0$.
 
The direction of this finite temperature current is opposite to the zero temperature edge current, meaning that the equilibrium edge current is reduced as temperature increases. 
 On the other hand, the total angular momentum carried by the system is a sum of two parts, the bulk (self-rotation) contribution $L_B(T)$ and the surface (orbital) contribution $\Delta L_S(T)\sim r_c\times j_S(T)$. The bulk contribution at low temperature is given by
\begin{eqnarray}
\label{bulkcurrent}
 \Delta L_B(T) & = & -{\Omega\over2}\frac{\pi r_c^2}{(2\pi)^2}\int_{0}^{\Lambda}d^2q (\hbar q)^2 n_B(\omega_q)  \\ \nonumber
 & \sim & -a\Omega\frac{r_c^2m^2(k_BT)^4}{4\hbar^2\rho_0^2U^2},
\end{eqnarray}
where $a=\int_{0}^{\infty} {x^3\over e^x-1}dx$.

In particular, we find from Eqs.\ (\ref{FiTcurrent}) and\ (\ref{bulkcurrent}) that the low temperature correction to total angular momentum
 $\Delta L(T)=L(T)-L(0)=\Delta L_B(T)+\Delta L_S(T)$ is dominated by the bulk (self-rotation) contribution at low temperature and total angular momentum of the system {\em increases} as function of temperature at low temperature although the edge current decreases as temperature raises.



\paragraph*{Experimental Implementation}
To realize our model experimentally, we need to (1) create pseudo-spin-1/2 bosons, (2) implement Rashba SOC, and (3) generate Zeeman splitting in the pseudo-spin space. The pseudo-spin 1/2 bosons can be created as in the experimental setup of Spielman's group\cite{lin2011spin} which breaks the degeneracy of $^{87}$Rb, $F=1$ ground state manifold and select the two internal spin states with lowest energy to be the basis of pseudo-spin-1/2 particles. Secondly, to generate 2D Rashba SOC, one may employ the proposal of Liu {\em et al.}\cite{liu2014realization} to apply two Raman fields induced by four lasers. The two Raman fields generate $(\sin{k_0 x}+i\sin{k_0 y})a^\dagger_{\uparrow} a_{\downarrow}$ term in the Hamiltonian which is 2D SOC. Thirdly, one needs to generate a {\em large} Zeeman energy splitting in this pseudospin basis. The small two-photon off resonance term mentioned in Ref.\cite{liu2014realization} may not be sufficient and another counter propagating Raman field is needed to realize Rabi oscillation as in Spielman's experiment\cite{lin2011spin}. Feshbach resonance can be applied to control the boson-boson interaction by tuning $s$-wave scattering length as a function of external magnetic field\cite{chin2010Feshbach}.

Last but not least, to detect the edge current in our system, one may conduct the time-of-flight(TOF) measurement for two spin species separated by Stern-Gerlach effect\cite{wang2012spin}. With SOC and interaction, we find a cross-like distribution around $\Gamma$ point when analyzing the density of minor component $n_{\downarrow}(\mathbf{k})$ (shown in Fig.\ref{Fig:densitydn}). The maxima at finite $k$ points are results of the edge mode.  
The chiral edge excitations can be detected directly by the Bragg scattering process which based on a two-photon process that directly transfers energy and momentum to an ensemble of atoms\cite{ernst2010probing}. Another possibility is to use quantum quench techniques\cite{killi2012use,killi2012anisotropic} which detects the structure of the boson wave function. One may apply a shaping potential well which is initially prepared to have a surrounding edge current. After removing this shaping potential, one can measure the cloud density revolution, revealing the information of angular momentum\cite{goldman2013direct}.
\begin{figure}
\centering
\includegraphics[width=3.5in]{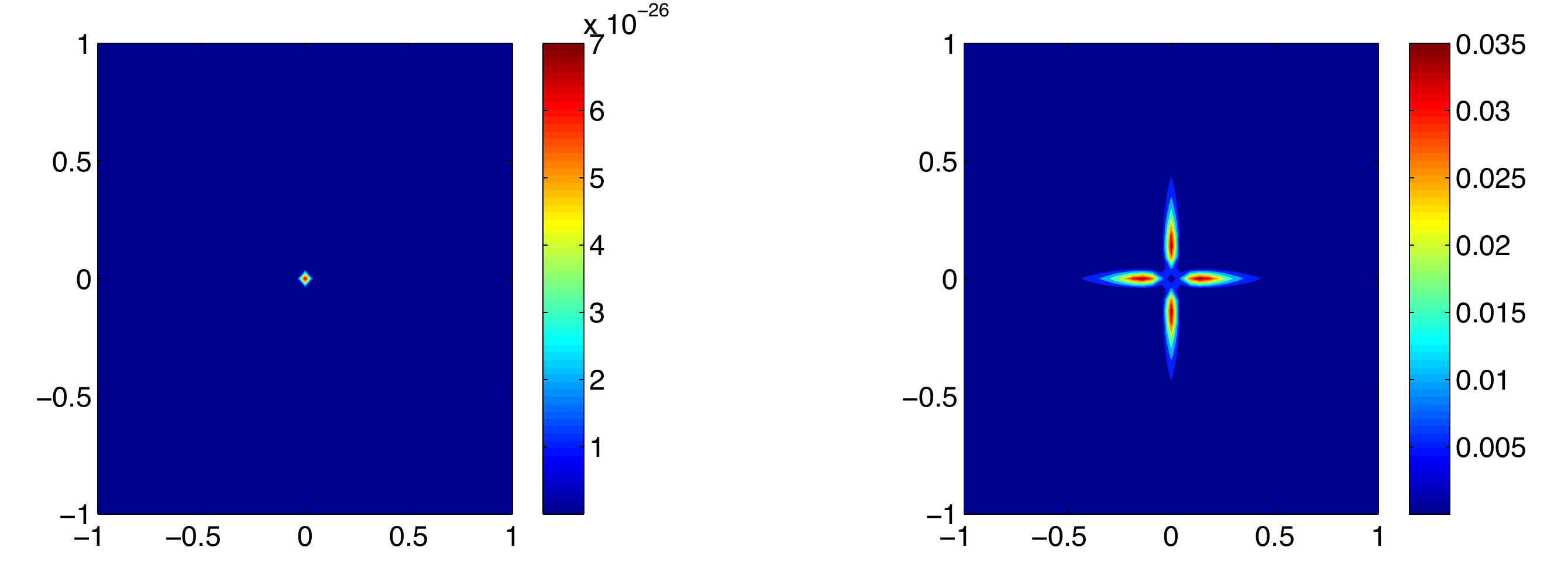}
\caption{The density distribution of the minor component in momentum space $n_{\downarrow}(\mathbf{k})$. Without SOC the density of the minor component at ground state is almost zero due to the Zeeman field splitting. With SOC and interaction $U=0.2$, the density distribution of $n_{\downarrow}(\mathbf{k})$ spreads out to higher momentum, with four maxima at finite $k$ points corresponding to edge mode. }\label{Fig:densitydn}
\end{figure}

Summarizing, we study in this paper the Gross-Pitaevskii (GP) equation for a simple model of weakly interacting bosons moving in a lattice with nonzero $k$-space Berry curvature at the momentum point where Bose-condensation occurs. By combining results from Local Equilibrium (Thomas Fermi) approximation and numerical self-consistent approach, we show that non-universal angular momentum and edge current exist in the system at zero temperature. The angular momentum has both orbital and self-rotation components. Furthermore, the equilibrium edge current is reduced by gapless chiral edge excitation as temperature increases but the total angular momentum carried by the system {\em increases} as temperature increases as a result of bulk self-rotation.

We acknowledge helpful discussion with Prof.Gyu Boong Jo and Prof. Jason Ho. This work is supported by HKRGC through grant No. HKUST/CEF/13G.

\bibliography{Berryboson}
%

\end{document}